\documentstyle[12pt,epic]{article}

\begin{document}

\begin{titlepage}
\title{On the integrability of $N$ = 2 supersymmetric massive theories}

\author{C\'esar G\'omez and Germ\'an Sierra
\\ Instituto de Matem\'aticas y F\'{\i}sica
Fundamental \\  CSIC, Serrano 123, E--28006 Spain.}

\date{ December 1993}

\maketitle

\begin{abstract}
In this paper we propose a criteria to establish the
integrability of
$N=2$ supersymmetric massive theories.
The basic data required are the vacua and the spectrum
of Bogomolnyi solitons, which can be neatly encoded in a graph
(nodes=vacua and links= Bogomolnyi solitons). Integrability
is then equivalent to the existence of solutions of a generalized
Yang-Baxter equation which is built up from the graph
( graph-Yang-Baxter equation).
We solve this equation for two
general types of graphs: circular and daisy,  proving, in
particular, the integrability of the following Landau-Ginzburg
superpotentials: $A_n$($t_1$), $A_n(t_2)$, $D_n$($\tau$),
$E_6$($t_7$), and  $E_8$($t_{16}$).
For circular graphs the solution are
intertwiners of the affine Hopf algebra
$\tilde{U}_{q}(A^{(1)}_1)$, while for daisy graphs the solution
corresponds to a susy generalization of the Boltzmann weights of
the chiral Potts model in the
trigonometric regime. A chiral Potts like solution is
conjectured for the more tricky case $D_n$$(t_2)$.
The scattering theory of circular models, for instance
$A_n$($t_1$) or $D_n$($\tau$),
is Toda like. The physical spectrum of daisy models,
as $A_n$($t_2$), $E_6$($t_7$) or $E_8$($t_{16}$),
is given by confined states of radial solitons. The
scattering theory of the confined states is again Toda like.
Bootstrap factors for the confined solitons are given by
fusing the susy chiral Potts $S-$matrices of the elementary
constituents; i.e the radial solitons of the daisy graph.

\end{abstract}

\end{titlepage}

\def\a{\alpha}
\def\l{\lambda}
\def\t{\theta}
\def\i{{\rm i}}
\def\s{ {\rm sinh}}
\def\c{ {\rm cosh}}
\def\x{ \frac{ \theta}{ 2 \pi {\rm i} } }

\section{Introduction and Summary}

It is a common believe among string practitioners that the space
of two dimensional theories will become, for strings, as
fundamental as space and time has been for classical physics
\cite{Za1}.
Independently whether this idea is true  or not, it is certainly a good
excuse for trying to unravel the secrets of two dimensional
models, in particular of those possessing $N=2$ supersymmetry. In
this paper we concentrate our efforts on the study of the
integrability and soliton scattering of two dimensional $N=2$
massive theories. The infinite symmetry of integrable models is
for off-shell strings the closest we can get to
on shell reparametrization invariance
(conformal symmetry) which fix the vacuum
solutions. We hope that the real dynamical meaning of integrability
will become clear at the end of the road, surely with the
arrival of a
well established string field theory.

Based on non renormalization theorems, $N=2$ massive theories
are usually described by a perturbed and non degenerate Landau-Ginzburg
superpotential \cite{Va1,Va2,Wa}.
In general a $N=2$ massive theory can
be described by a graph with the nodes representing the vacua and
the links Bogomolnyi solitons \cite{Ho}.  The critical values are then
interpreted as coordinates for the nodes and the soliton fermion
numbers as labeling the links. An interesting question
consist in deriving this graph, for susy preserving
deformations of $N=2$ superconformal field theories, directly
from its chiral ring structure \cite{Va2}. Notice that the graph
contains information on the spectrum of the theory.

The first problem we address in this paper is the integrability
of $N=2$ massive theories characterized by the graph of vacua
and Bogomolnyi solitons. We translate the question of
integrability, i.e. existence of an infinite number of conserved
charges, into the existence of solutions
of a generalization of the
standard Yang-Baxter equation
based on the graph. We denote this new equation graph
-Yang-Baxter (gYB) \cite{GS} . We will consider two generic types of
graphs characterized by their shape :
circular with all the nodes living on the same circle and
daisy with all the nodes on a circle except one located at its
center. In both cases we find solutions to the corresponding gYB
equation provided we restrict the two-soliton scattering
processes to those which are elastic and with no interchange of
fermion number. The solutions for circular graphs have a neat quantum
group meaning. In fact they are the intertwiners of the affine
Hopf algebra
$U_q( A^{(1)}_1)$ extended with to extra central elements.
This extension is denoted
in the literature $\tilde{U}_q( A^{(1)}_1)$ \cite{Ji}. In
this way we are able, for instance, to prove the integrability
of the $D_n (\tau)$-perturbation which was unknown until now \cite{LW} .
 The solution for this case only differs from the well known
$A_n(t_1)$  $S$-matrix
\cite{Fe1,Fe2} by the eigenvalues
of the extra central elements. For daisy graphs the solution, as
advertized in reference \cite{GS}, is given by a $N=2$ extension
of chiral Potts at the superintegrable point. This solution
allows us to prove the integrability of models like $E_6$$(t_7)$
and $E_8(t_{16})$.
All these examples, namely  $D_n(\tau), E_6(t_7)$
and $E_8(t_{16})$, appear
with questions marks in the classification given in reference \cite{Ga}.
The explicit proof of integrability presented in this work support
the phenomenological observations of \cite{Ga}. For the $D_n(t_2)$ we
conjecture a solution which combines general chiral Potts and
the Ising model.

The second problem is the definition of a scattering theory for
solitons or in other words, to find solutions of the gYB
equation satisfying crossing, unitarity and the bootstrap
relations \cite{ZZ}. For circular graphs we don't find any surprise, the
physical spectrum is the one defined by the graph with the bootstrap
structure reflecting its geometry. The scattering is  Toda
like \cite{Toda,To1,To2}.
The surprise and the possible new physics come with daisy
graphs. In this case there are not  crossing and unitary
solutions of the gYB equation. We interpret this result as
indicating that the radial solitons and antisolitons don't
define the physical spectrum of the theory. The bound states of
radial solitons are living on a circular graph and we can expect
their $S$-matrix, which can be obtained by fusion of the
$S$-matrices for the radial solitons, to be a solution of the
circular gYB equation. This in fact the case, moreover the
solution is an intertwiner of $\tilde{U}_q(A^{(1)}_1)$ with non trivial
eigenvalues for the extra central elements. From this result we
conclude that the physical spectrum of daisy models is a
"confined" spectrum defined by the subset of nodes on the
circle. This phenomena is in a certain sense dual to the one
described in \cite{Smirnov} for the soliton free regime in
sine-Gordon model. There, the soliton free spectrum describes non
unitary theories, while here we need to confine the spectrum in
order to get a physical $S$-matrix satisfying bootstrap.
For the Landau-Ginzburg superpotential $W$= $x^{k+2}$-$tx^2$ the
previous result implies that their physical, confined
spectrum is, up to fermion numbers, the same one gets for the
circular Landau-Ginzburg model $W$= $x^{k+1}$-$tx$. The circular
model is the most relevant perturbation of the minimal SCFT with
central extension $c= \frac{3 (k-1)}{k +1}$
while the one we start with is the
next to the most relevant perturbation of the SCFT with central
extension $c=  \frac{3 k}{k +2}$ An important issue is therefore the
ultraviolet behaviour of daisy potentials or in other words whether
the infrared "confinement" is or not associated with asymptotic
freedom in the ultraviolet.

The plan of the paper is as follows. In section II we give the
general definitions and explain the integrability criteria of
$N=2$ massive theories. In section III we study circular and
daisy graphs proving their integrability and aplying the results
to some Landau-Ginzburg models whose integrability was unclear.
In section IV we consider the scattering theory and discuss the
confined spectrum for daisy graphs.

\section{N=2 massive theories}
\subsection{Definitions}

Let us denote by ${\cal A}$ the $N=2$ algebra:

\begin{eqnarray}
( Q^{\pm})^2=( \bar{Q}^{\pm})^2= &
\{Q^+, \bar{Q}^- \}=  \{Q^-,
 \bar{Q}^+ \}= & 0 \;, \nonumber \\
\{ Q^+,{Q}^- \}= P, & \{ \bar{Q}^+,
 \bar{Q}^- \}= \bar{P} \;, &
\label{ 1 } \\
\{ Q^+, \bar{Q}^+ \} = W,  &
\{ {Q}^-, \bar{Q}^- \}= \bar{W} \;, & \nonumber \\
\left[ {\cal F},Q^{\pm} \right]=\pm Q^{\pm}, &
\left[ {\cal F},\bar{Q}^{\pm} \right]=\mp \bar{Q}^{\pm}
& \nonumber
\end{eqnarray}

\noindent
where $Q^{\pm}$,$\bar{Q}^{\pm}$ are the susy generators, $W$ the
topological central term  \cite{WittenO} and $\cal F$ the fermion number.
 The
$N=2$ algebra is equipped with a coalgebra structure defined by
the following comultiplication rules:

\begin{eqnarray}
\Delta Q^{\pm} & = & Q^{\pm} \otimes {\bf 1} +
e^{ \pm {\rm i} \pi \cal{F} } \otimes Q^{\pm} \nonumber \\
\Delta \bar{Q}^{\pm} & = & \bar{Q}^{\pm} \otimes {\bf 1} +
e^{ \mp {\rm i} \pi \cal{F} } \otimes \bar{Q}^{\pm}
\label{2} \\
\Delta W & = & W \otimes {\bf 1} + {\bf 1} \otimes W \nonumber \\
\Delta P & = & P \otimes {\bf 1} + {\bf 1} \otimes P \nonumber
\end{eqnarray}

By a massive theory we mean one with a mass gap and non
degenerate vacua. Given a couple of ordered vacua $(i,j)$ a
soliton $ s_{i,j} $ is a field configuration satisfying the
equation of motion and
which connects the vacua $i$ at $x= - \infty$ with
the vacua  $j$ at $x= \infty$. Following the
spirit of reference \cite{Ho} we define a massive
$N=2$ theory by the following set of data:

 {\bf D1)} A graph $\cal{G}$ characterized by a set of
nodes $ {i,j,..}$ and positive integer numbers $ \mu_{i, j}$
which counts the
number of links between the nodes $i$ and $j$.

 {\bf D2)} Complex coordinates $ w_{i} $ for the nodes and real
numbers $ f_{i, j} $ associated with each ordered couple of
connected nodes.

The physical interpretation of these data is as follows. The
nodes will represent the different vacua configurations. Each
ordered couple of connected nodes is interpreted as a Bogomolnyi
soliton supermultiplet,
transforming under the $ N=2 $ supersymmetry as follows:

\begin{eqnarray}
\pi_{i,j}(\theta)(Q^-) = \left( \begin{array}{cc}
0 & 0 \\ \sqrt{m_{i, j}} e^{\theta/2}  & 0 \end{array} \right),&
\pi_{i,j}(\theta)(Q^+)= \left( \begin{array}{cc}
0 &  \sqrt{m_{i, j}} e^{\theta/2} \\ 0 & 0  \end{array} \right)&
\nonumber \\
\pi_{i,j}(\theta)(\bar{Q}^+) = \left( \begin{array}{cc}
0 & 0 \\ \omega_{i, j}  \sqrt{m_{i, j}}
e^{-\theta/2}  & 0 \end{array} \right),&
\pi_{i,j}(\theta)(\bar{Q}^-)= \left( \begin{array}{cc}
0 & \omega_{i, j}^*  \sqrt{m_{i, j}}
e^{-\theta/2} \\ 0 & 0  \end{array} \right)
& \label{3}   \\
\pi_{i, j}(\theta)({\cal F}) =
\left( \begin{array}{cc} f_{i, j} & 0 \\ 0 & f_{i, j} -1
\end{array} \right)&  & \nonumber
\end{eqnarray}

\noindent
where:
\begin{eqnarray}
m_{i, j} = 2 | \Delta_{i, j} |, & \omega_{i, j} = \frac{
\Delta_{i, j} }{ | \Delta_{i, j} | }, & \Delta_{i, j}= w_j -w_i
\label{4}
\end{eqnarray}
The parameters $\theta$, $ m_{i, j} $, $ \Delta_{i, j} $ and $
f_{i, j}$
are respectively the rapidity, mass, topological and
fermion number of
the Bogomolnyi soliton $s_{i, j}$.

The $ N=2$ massive theory admits a Landau-Ginzburg
interpretation if there exist a complex superpotential $W$
whose critical points are in correspondence with the
$N=2$ vacua and  such
that the complex coordinates $w_{i}$, introduced above, coincide
with the critical values of $W$, and the fermion numbers
$f_{i,j}$ are given by the index theorem formula:

\begin{equation}
{\rm exp}( 2 \pi {\rm i} f_{j, k})
= {\rm phase} \left( \frac{ {\rm det} H(k)}{ {\rm det} H(j)}
\right)
\label{5}
\end{equation}

\noindent
with $H(j)$ the Hessian of $W$ at the critical point $j$.
Eq.( \ref{5}) fixes the fermion number only modulo 1. In
table 1 we collect the basic information concerning
some simple Landau-Ginzburg superpotentials.

\begin{table}
\begin{center}
\begin{tabular}{|c|c|c|c|c|}
\hline
Model & Superpotential W & Extrema &W-values &
$f_{i,j}$  \\ \hline
$A_{k+1}(t_1)$ &$ \frac{x^{k+2}}{k+2} -x$ &
$\begin{array}{l}  x_j = e^{ \frac{ 2 \pi {\rm i} j}{k+2}} \\ j=1,\dots
, k+1  \end{array}$ &$ W_j =e^{\frac{2 \pi {\rm i} j}{k+1}}$
& $f_{j,j+1}=\frac{1}{k+1}$  \\ \hline
$A_{k+1}(t_2)$ &$ \frac{x^{k+2}}{k+2} -\frac{x^2}{2}$ &
$\begin{array}{l} x_* = 0 \\
x_j = e^{ \frac{ 2 \pi {\rm i} j}{k}} \\ j=1,\dots
, k \end{array}$ & $\begin{array}{l} W_* =0 \\
W_j = e^{ \frac{ 4 \pi {\rm i} j}{k}} \end{array}$
& $\begin{array}{l} f_{*,j}= \frac{1}{2} \\
f_{j,*}= \frac{1}{2} \end{array}$ \\ \hline
$A_{k+1}(t_k) $ & $ \frac{T_{k+2}(x)}{k+2}$ &
$\begin{array}{l} x_j = 2 {\rm cos}\left( \frac{\pi j}{k+2}
\right) \\ j = 1, \cdots ,k+1 \end{array} $
& $W_j = (-1)^j $ & $f_{j,j+1}= \frac{1}{2}$  \\ \hline
$D_{k+3}(\tau)$ &$ \frac{x^{k+2}}{2(k+2)} +\frac{x y^2}{2} -y$
&$ \begin{array}{l} (x_j, y_j)= (e^{ \frac{2 \pi {\rm i} j}{k+3}},
e^{- \frac{2 \pi {\rm i} j}{k+3}}) \\ j= 1,\cdots, k+3 \end{array}$
& $W_j= e^{- \frac{2 \pi {\rm i} j}{k +3}} $&  $f_{j,j+1}=\frac{2}{k+3}$
\\ \hline
$D_{k+2}(t_2)$ &
$ \frac{x^{k+1}}{2(k+1)} + \frac{x y^2}{2} - x $ & $ \begin{array}{l}
(x_{*_1}, y_{*_1})= (0,\sqrt{2}) \\
(x_{*_2}, y_{*_2})= (0,- \sqrt{2})
\\ (x_j,y_j)= (2^{1/k} e^{ \frac{2 \pi {\rm i} j}{k} } ,0)
\\ j=1,\dots, k \end{array} $ &
$\begin{array}{l} W_{*_{1,2}} = 0 \\
W_j = e^{ \frac{2 \pi {\rm i} j}{k}} \end{array} $ &
$\begin{array}{l} f_{{*_{1,2}},j}= \frac{1}{2} \\
f_{j,{*_{1,2}}}= \frac{1}{2} \end{array}$ \\ \hline
$E_6(t_7)$ & $ \frac{x^3}{3} + \frac{y^4}{4} - x y$ & $ \begin{array}{l}
(x_*, y_*)= (0,0) \\ (x_j,y_j)= (e^{ \frac{2 \pi {\rm i} j}{5}},
e^{ \frac{4 \pi {\rm i} j}{5}}) \\ j=1,\dots, 5 \end{array} $ &
$\begin{array}{l} W_* = 0 \\
W_j = e^{ \frac{6 \pi {\rm i} j}{5}} \end{array} $ &
$\begin{array}{l} f_{*,j}= \frac{1}{2} \\
f_{j,*}= \frac{1}{2} \end{array}$ \\ \hline
$E_8(t_{16})$ & $ \frac{x^3}{3} + \frac{y^5}{5} - x y$ & $ \begin{array}{l}
(x_*, y_*)= (0,0) \\ (x_j,y_j)= (e^{ \frac{2 \pi {\rm i} j}{7}},
e^{ \frac{4 \pi {\rm i} j}{7}}) \\ j=1,\dots, 7 \end{array} $ &
$\begin{array}{l} W_* = 0 \\
W_j = e^{ \frac{6 \pi {\rm i} j}{7}} \end{array}  $ &
$\begin{array}{l} f_{*,j}= \frac{1}{2} \\
f_{j,*}= \frac{1}{2} \end{array}$ \\ \hline
\end{tabular}
\caption{All these examples are perturbations of ADE Landau-Ginzburg
models.The superpotential $T_{k+2}(x)$ of the model $A_{k+1}$
is the Chebishev polynomial $T_n( 2 \; {\rm cos} \t
) = 2 \; {\rm cos} \; n \t$.}
\end{center}
\label{1}
\end{table}

\subsection{ Integrability conditions }

In this section we propose an integrability criteria for $N=2$
massive theories. To characterize these theories we will use the
geometrical data D1) and D2) defined in the previous section.
As
a preliminary step we first introduce some notational background.
Given the $N=2$ graph $\cal{G}$ a plaquette
$\left(
\begin{array}{@{\,}c@{\,}c@{\,}} i & l \\ j & k \end{array} \left|
\begin{array}{@{\,}c@{\,}c@{\,}}
\a_3 & \a_4 \\ \a_1 & \a_2 \end{array} \right) \right. $
is defined by a
set of four connected nodes ${i,j,k,l}$ and four extra labels
${\a_1,\dots,\a_4}$ which we introduce in order to differenciate
solitons interpolating the same two vacua,  i.e.  $ \a_1=1,2,..,
\mu_{i, j}$,
etc (see figure 1).

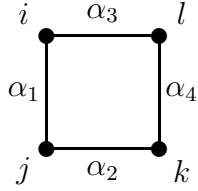
\begin{figure}
\begin{center}

\unitlength = 0.75mm
\begin{picture}(40,40)(-10,-10)
\drawline(0,20)(20,20)(20,0)(0,0)(0,20)
\put(0,0){\circle*{3}}
\put(0,20){\circle*{3}}
\put(20,20){\circle*{3}}
\put(20,0){\circle*{3}}
\put(-4,24){\makebox(0,0){$i$}}
\put(-4,-4){\makebox(0,0){$j$}}
\put(24,-4){\makebox(0,0){$k$}}
\put(24,24){\makebox(0,0){$l$}}
\put(-4,10){\makebox(0,0){$ \alpha_1$}}
\put(10,-4){\makebox(0,0){$ \alpha_2$}}
\put(10,24){\makebox(0,0){$ \alpha_3$}}
\put(24,10){\makebox(0,0){$ \alpha_4$}}
\thicklines
\end{picture}

\end{center}
\caption[]{ Plaquette}
\end{figure}

$N=2$-invariant plaquettes are those satisfying:

\begin{eqnarray}
\pi_{i,j}\otimes \pi_{j,k}(\Delta(z)) &=& \pi_{i,l} \otimes
\pi_{l,k}(\Delta(z))
\label{6}
\end{eqnarray}

\noindent
for $z$ any central element of the $N=2$ algebra and the irreps
the ones defined in equation (\ref{3}) . We shall say that a $N=2$ invariant
plaquette is elastic if it satisfies the following extra
condition for the masses:

\begin{eqnarray}
m_{i, j} = m_{l, k} &, & m_{j, k} = m_{i, l} \label{7}
\end{eqnarray}

Each elastic plaquette is associated with a four by four
elastic $S-$matrix
(see figure 2): $S^{m_3 \; m_{4}}_{m_{1} \; m_{2}} \left(
\begin{array}{@{\,}c@{\,}c@{\,}} i  & l   \\ j   & k   \end{array} \left|
\begin{array}{@{\,}c@{\,}c@{\,}} \a_3  & \a_4   \\ \a_1   & \a_2   \end{array}
\right) \right. (\theta)$ ,which describes the transition
amplitude of the scattering process
$s_{i j}(\t_1) + s_{j k}(\t_2) \rightarrow
s_{i l}(\t_2) + s_{l k}(\t_1)$. The
N=2 labels $m_1,\dots ,m_4$ take two values $ 0$
and $1$ ,recall eq. (\ref{3}), and refer to the so called
up(u) and down(d) Bogomolnyi states.

After these definitions we propose the following integrability
criteria for $N=2$ massive theories.

A $N=2$ massive theory is integrable if
the following two conditions hold:

${\bf C_1}$: $N=2$ Invariance.
The  $S$-matrix associated to each elastic plaquette
is an intertwiner relative to
the coalgebra structure of the $N=2$ algebra ${\cal A}$:

\begin{eqnarray}
& S\left( \begin{array}{@{\,}c@{\,}c@{\,}} i  & l  \\ j   & k  \end{array}
\left|
\begin{array}{@{\,}c@{\,}c@{\,}}  \a_3  & \a_4   \\ \a_1   & \a_2 \end{array}
\right) \right. (\theta_{12})
( \pi_{i,j}(\theta_1) \otimes \pi_{j,k}(\theta_2) )
\Delta (g)  =& \label{8} \\
& ( \pi_{i,l}(\theta_2) \otimes
\pi_{l,k}(\theta_1) ) \Delta (g)
S\left( \begin{array}{@{\,}c@{\,}c@{\,}} i  & l  \\ j   & k  \end{array} \left|
\begin{array}{@{\,}c@{\,}c@{\,}}  \a_3  & \a_4   \\ \a_1   & \a_2  \end{array}
\right) \right. (\theta_{12})&
\nonumber
\end{eqnarray}

\noindent
for g any element of the $N=2$ algebra.

\begin{figure}
\begin{center}

\unitlength = .75mm
\begin{picture}(90,90)(0,0)
\drawline(30,30)(60,30)(60,60)(30,60)(30,30)
\put(30,30){\circle*{3}}
\put(60,30){\circle*{3}}
\put(60,60){\circle*{3}}
\put(30,60){\circle*{3}}
\put(26,64){\makebox(0,0){$i$}}
\put(26,26){\makebox(0,0){$j$}}
\put(64,26){\makebox(0,0){$k$}}
\put(64,64){\makebox(0,0){$l$}}
\put(26,50){\makebox(0,0){$\alpha_1$}}
\put(38,26){\makebox(0,0){$\alpha_2$}}
\put(38,64){\makebox(0,0){$\alpha_3$}}
\put(64,50){\makebox(0,0){$\alpha_4$}}
\thicklines
\dottedline{1.4}(45,15)(45,75)
\dottedline{1.4}(15,45)(75,45)
\put(8,45){\makebox(0,0){$\theta_1$}}
\put(45,7){\makebox(0,0){$\theta_2$}}
\put(22.5,41){\makebox(0,0){$m_1$}}
\put(67.5,41){\makebox(0,0){$m_4$}}
\put(49,22.5){\makebox(0,0){$m_2$}}
\put(49,67.5){\makebox(0,0){$m_3$}}
\drawline(43.5,73.5)(45,75)(46.5,73.5)
\drawline(73.5,46.5)(75,45)(73.5,43.5)

\end{picture}

\end{center}
\caption[]{$ N=2 \; $ S-matrix}
\end{figure}
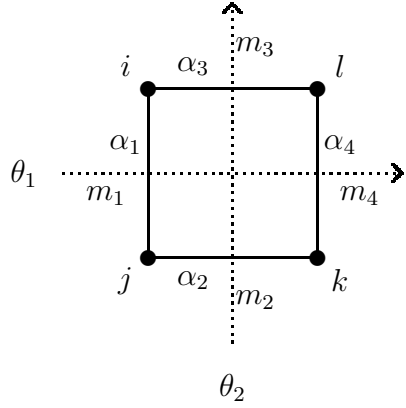

${\bf C_2}$: Graph-Yang-Baxter Equation (gYB):

\begin{eqnarray}
\sum_{p, m',\a' }
S^{m'_1 m'_2}_{m_1 m_2} \left( \begin{array}{@{\,}c@{\,}c@{\,}}
i & p \\ j & k \end{array}
\left|
\begin{array}{@{\,}c@{\,}c@{\,}} \a'_1 & \a'_2 \\ \a_1 &  \a_2 \end{array}
\right) \right.
(\theta) \;
S^{m'_3 m''_3}_{m'_2 m_3} \left( \begin{array}{@{\,}c@{\,}c@{\,}}
p & r \\ k & l \end{array}
\left|
\begin{array}{@{\,}c@{\,}c@{\,}} \a'_3 & \a''_3 \\ \a'_2 & \a_3 \end{array}
\right) \right. (\theta+ \theta' ) \;
S^{m''_1 m''_2}_{m'_1 m'_3} \left( \begin{array}{@{\,}c@{\,}c@{\,}}
i & s \\ p & r \end{array}
\left|
\begin{array}{@{\,}c@{\,}c@{\,}} \a''_1 & \a''_2 \\ \a'_1 & \a'_3
\end{array} \right)
\right. (\theta')  & &  \nonumber \\
&  &
\label{9} \\
= \sum_{p, m',\a' }
S^{m'_2 m'_3}_{m_2 m_3} \left( \begin{array}{@{\,}c@{\,}c@{\,}}
j & p \\ k & l \end{array}
\left|
\begin{array}{@{\,}c@{\,}c@{\,}} \a'_2 & \a'_3 \\ \a_2 & \a_3 \end{array}
\right)
\right. (\theta') \;
S^{m''_1 m'_1}_{m_1 m'_2} \left( \begin{array}{@{\,}c@{\,}c@{\,}}
i & s \\ j & p \end{array}
\left|
\begin{array}{@{\,}c@{\,}c@{\,}} \a''_1 & \a'_1 \\ \a_1& \a'_2 \end{array}
\right)
\right. (\theta+ \theta' ) \;
S^{m''_2 m''_3}_{m'_1 m'_3} \left( \begin{array}{@{\,}c@{\,}c@{\,}}
s & r \\ p & l \end{array}
\left|
\begin{array}{@{\,}c@{\,}c@{\,}} \a''_2 & \a''_3 \\
\a'_1& \a'_3 \end{array} \right)
\right. (\theta)  & &
\nonumber
\end{eqnarray}

\noindent
where the sum over graph labels is restricted to elastic plaquettes
(see figure 3).Observe that the gYB equation is a "fusion" of
the ordinary vertex YB equation for the labels $m$ and $\a$ and
the RSOS YB equation for the labels $i,j,etc.$.

\begin{figure}
\begin{center}

\unitlength = 1mm
\begin{picture}(140,60)(-30,-30)
\drawline(0,0)(10,17.32)(-10,17.32)(-20,0)(0,0)(10,-17.32)
(-10,-17.32)(-20,0)
\drawline(10,17.32)(20,0)(10,-17.32)
\put(0,0){\circle*{2}}
\put(10,17.32){\circle*{2}}
\put(-10,17.32){\circle*{2}}
\put(-20,0){\circle*{2}}
\put(-10,-17.32){\circle*{2}}
\put(10,-17.32){\circle*{2}}
\put(20,0){\circle*{2}}
\thicklines
\drawline(80,0)(100,0)(90,17.32)(70,17.32)(80,0)(70,-17.32)
(90,-17.32)(100,0)
\drawline(70,17.32)(60,0)(70,-17.32)
\put(80,0){\circle*{2}}
\put(100,0){\circle*{2}}
\put(90,17.32){\circle*{2}}
\put(70,17.32){\circle*{2}}
\put(60,0){\circle*{2}}
\put(70,-17.32){\circle*{2}}
\put(90,-17.32){\circle*{2}}
\put(-24,0){\makebox(0,0){$i$}}
\put(56,0){\makebox(0,0){$i$}}
\put(-15,-20){\makebox(0,0){$j$}}
\put(65,-20){\makebox(0,0){$j$}}
\put(-15,20){\makebox(0,0){$s$}}
\put(65,20){\makebox(0,0){$s$}}
\put(-3,2){\makebox(0,0){$p$}}
\put(15,-20){\makebox(0,0){$k$}}
\put(95,-20){\makebox(0,0){$k$}}
\put(24,0){\makebox(0,0){$l$}}
\put(104,0){\makebox(0,0){$l$}}
\put(15,20){\makebox(0,0){$r$}}
\put(95,20){\makebox(0,0){$r$}}
\put(82,2){\makebox(0,0){$p$}}
\put(32,0){$=$}
\put(-35,0){\makebox(0,0){$\sum_{p,\{ m' \}}$}}
\put(45,0){\makebox(0,0){$\sum_{p,\{ m' \}}$}}
\dottedline{1.4}(-5,25)(-5,-25)
\dottedline{1.4}(85,25)(85,-25)
\dottedline{1.4}(-20,-17.5)(26,9.2)
\dottedline{1.4}(54,-9.2)(100,17.3)
\dottedline{1.4}(54,9.2)(100,-17.3)
\dottedline{1.4}(-20,17.3)(26,-9.2)
\put(-24,-19.3){\makebox(0,0){$m_1$}}
\put(-5,-28){\makebox(0,0){$m_2$}}
\put(29.3,-11){\makebox(0,0){$m_3$}}
\put(29.3,11){\makebox(0,0){$m''_3$}}
\put(-5,28){\makebox(0,0){$m''_2$}}
\put(-24,19.3){\makebox(0,0){$m''_1$}}
\put(-9,4){\makebox(0,0){$m'_1$}}
\put(-3,-11){\makebox(0,0){$m'_2$}}
\put(6,4){\makebox(0,0){$m'_3$}}
\put(50,-10){\makebox(0,0){$m_1$}}
\put(85,-28){\makebox(0,0){$m_2$}}
\put(102,-20){\makebox(0,0){$m_3$}}
\put(50,10){\makebox(0,0){$m''_1$}}
\put(85,28){\makebox(0,0){$m''_2$}}
\put(102,20){\makebox(0,0){$m''_3$}}
\put(74,4){\makebox(0,0){$m'_1$}}
\put(81,-10){\makebox(0,0){$m'_2$}}
\put(88,4){\makebox(0,0){$m'_3$}}
\end{picture}

\end{center}
\caption[]{Graph-Yang-Baxter equation. We have not included the
$\alpha's$ labels in order to simplify the drawing.}
\end{figure}
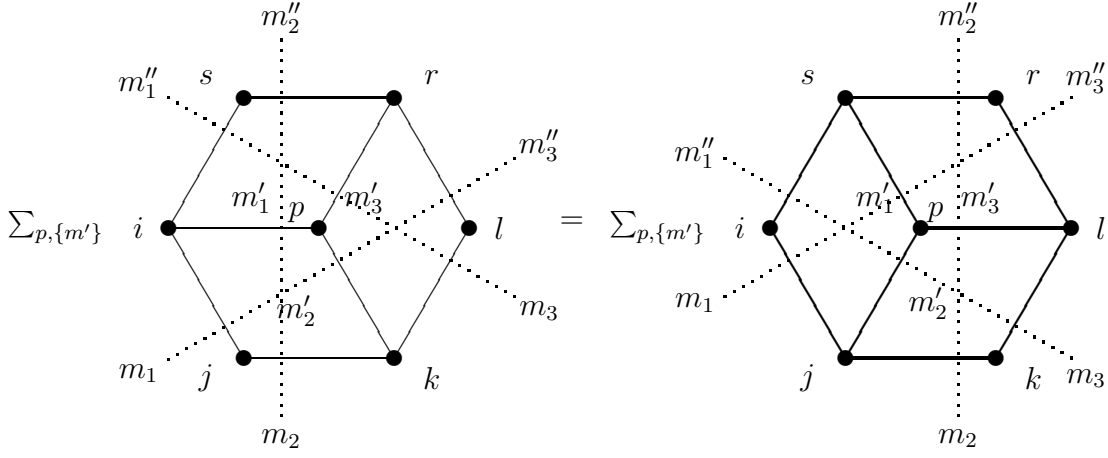

We shall make now some comments:

1) Using the fact that different solitons interpolating the
same two vacua are associated with the same $N=2$ irrep we obtain
the following factorization of the $S$-matrix:

\begin{eqnarray}
S^{m_3 \; m_{4}}_{m_{1} \; m_{2}} \left(
\begin{array}{@{\,}c@{\,}c@{\,}} i  & l   \\ j   & k   \end{array} \left|
\begin{array}{@{\,}c@{\,}c@{\,}} \a_3  & \a_4   \\ \a_1   & \a_2   \end{array}
\right) \right. (\theta) \;\;
=S^{(N=2)\;  m_3 \; m_{4}}_{\;\;\;\;\;\;\;\;\;\;\;\; m_{1} \; m_{2}}
\left(
\begin{array}{@{\,}c@{\,}c@{\,}} i  & l   \\ j   & k   \end{array}
\right)(\theta)
S^{(N=0)} \left(
\begin{array}{@{\,}c@{\,}c@{\,}} i  & l   \\ j   & k   \end{array} \left|
\begin{array}{@{\,}c@{\,}c@{\,}} \a_3  & \a_4   \\ \a_1   & \a_2  \end{array}
\right) \right. (\theta)  & &
\label{10}
\end{eqnarray}

  Notice that the requirement $C_{1}$ of $N=2$ invariance, fix the ratios
for the entries of the $N=2$ piece of the $S$-matrix leaving
completely undetermined the $N=0$ part. The simplest example of
the factorization (\ref{10}) takes place for the $ADE$
Chebishev potentials which describe the least relevant
perturbation of minimal SCFT. In these models the $N=2$ piece is
independent of the vacua labels and the $N=0$ piece is a
solution to the standard $RSOS$ Yang Baxter equation for the
$ABF$ models \cite{ABF} defined by the Coxeter $ADE$
graphs \cite{Fe2} .  When the $w$-coordinates are
degenerate, i.e. two o more nodes of the graph correspond to the
same point in the $w$-plane, then different links of the graph
will be mapped into the same one in the $w$-plane. All these links
are associated with the same $N=2$ irrep. Therefore the $N=2$
part of the $S$ matrix only depends on the $w$-plane graph.
The model is completely characterized by the $N=2$ part if the
multiplicities $\mu_{i,j}$ are all equal to one and
the $w$-coordinates are non degenerate, in all other cases there exist
a $N=0$ part.

2) If all multiplicities $\mu_{i,j}$ are equal to one, then the
gYB equation (\ref{9})
for the choices $m_1=m_2=m_3=m_4=0 $ or $1$ reduces
to a RSOS Yang-Baxter equation for the graph $\cal{G}$ with the
extra selection rule imposed by the elasticity condition
(\ref{7}). Solutions to equation (\ref{8}) and to the full
fledged gYB equation  (\ref{9}) would be equivalent, in these
cases, to the existence of a $N=2$ supersymmetric extension of
the corresponding $RSOS$ statistical model.

3) The $N=2$ massive theories can possess, in some particular cases,
extra $N=0$ symmetries that would impose additional
restrictions on the $N=0$ part of the $S$-matrix. The $N=0$
symmetry, if present, will act on the multiplicity labels
($\a$)  fixing the ratios for the different entries of the
$N=0$ $S$-matrix. This $N=0$ symmetry can be associated with a
classical or a quantum algebra and in all cases commute with the
$N=2$ supersymmetry.  $CP^n$ sigma models and affine perturbations
of Kazama-Suzuki cosets are good examples of $N=2$ massive
theories with $N=0$ symmetries.

4)  Extra integrability symmetries. A possible scenario  we
will often find  is that of $N=2$ massive theories where
the gYB equation has not solution, unless we restrict the set of
elastic plaquettes. Of course this restriction should not imply
the restriction of the graph, i.e. the physical decoupling
of any vacua. If this is the case
we would conclude that the theory is integrable
and that the extra selection rule is an integrability symmetry.
In all the integrable $N=2$ massive theories we have studied,
integrability is obtained after reducing the elastic plaquettes
to those with equal fermion numbers for opposite sides and
therefore the fermion number appears as an "individual"
conserved quantity.
In principle, this extra selection rule is independent of
the general $Z$-invariance \cite{Bax} which underlines integrability.

5) The integrability of the $N=2$ massive theories, characterized
by conditions $C_1$ and $C_2$ should not be
confussed with the existence of a well defined scattering
theory. As we will discuss in section IV in order to define a
scattering theory we need to impose the additional physical
requirements of unitarity, crossing and bootstrap.

6) The geometrical data $D_1$ and $D_2$ we have used to define a
massive $N=2$ theory, together with the
conditions $C_1$ and $C_2$ in terms of which we characterize its
integrability, can be interpreted from a mathematical point of
view, as a way to define a new mathematical structure which
generalizes that of quantum groups. This new mathematical object
or, graph quantum group, would be defined by a Hopf algebra $\cal{A}$, a
graph, and a map from links of the graph into irreps of the Hopf
algebra in such a way that for any plaquette of the graph there
exist an intertwiner satisfying conditions $C_1$ and $C_2$. The
main difference with respect to ordinary quantum groups is that now
the intertwiner stablish an equivalence between irreps which are
not necessarily related by a permutation. Therefore these
intertwiners cannot follow from a universal $R$-matrix.

\section{ Partial Classification of Integrable $N=2$ Massive Theories}

In this section we present a partial classification of
integrable $N=2$ theories, mostly based on the geometry of the
graph. We consider two basic types of geometries, which seem to
be the basic building blocks of most of the known soliton
polytopes \cite{To2} , namely circular graphs with all the nodes on the same
circle and daisy graphs with all nodes on the same circle except
one which is located at the center (see fig. 4).

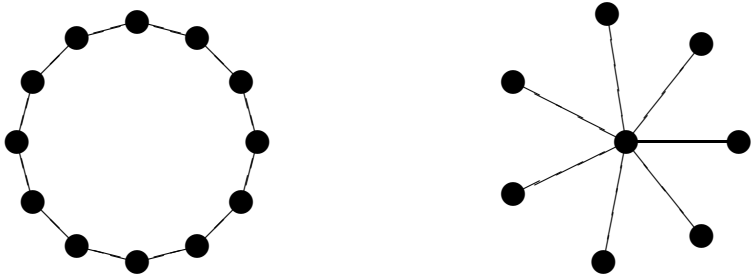
\begin{figure}
\begin{center}

\unitlength = 1mm
\begin{picture}(110,45)(0,0)
\drawline(105,25)(90,25)(100,38)
\drawline(87.5,42)(90,25)(75,33)
\drawline(75,18)(90,25)(87,9)
\drawline(90,25)(100,12.5)
\put(90,25){\circle*{3}}
\put(105,25){\circle*{3}}
\put(100,38){\circle*{3}}
\put(87.5,42){\circle*{3}}
\put(75,33){\circle*{3}}
\put(75,18){\circle*{3}}
\put(87,9){\circle*{3}}
\put(100,12.5){\circle*{3}}
\drawline(41,25)(38.85,33)(33,38.85)(25,41)(17,38.85)(11.15,33)
(9,25)(11.15,17)(17,11.15)(25,9)(33,11.15)(38.85,17)(41,25)
\put(41,25){\circle*{3}}
\put(38.85,33){\circle*{3}}
\put(33,38.85){\circle*{3}}
\put(25,41){\circle*{3}}
\put(17,38.85){\circle*{3}}
\put(11.15,33){\circle*{3}}
\put(9,25){\circle*{3}}
\put(11.15,17){\circle*{3}}
\put(17,11.15){\circle*{3}}
\put(25,9){\circle*{3}}
\put(33,11.15){\circle*{3}}
\put(38.85,17){\circle*{3}}
\end{picture}

\end{center}
\caption[]{Circular and daisy graphs}
\end{figure}

\subsection{ Circular graphs}

Let be a generic circular graph with $n$ uniformally distributed
nodes $j=1,2,\dots,n$ and links $(j,j+r)$ connecting arbitrary
couples of different nodes. The simplest links $(j,j+1)$ and $(j,j-1)$ will be
interpreted as the fundamental soliton and antisoliton. To fix the model we
give the fermion number $f$  of the fundamental
solitons and the $w$-coordinates of the nodes. Without loss of
generality we can take as $w$-coordinates the circular angles
as described in figure 5. The
fermion number of non fundamental solitons $(j,j+r)$ is
determined by the comultiplication rule of the fermion
number. Notice that for circular graphs the fermion number of the
antisoliton $(j,j-1)$ coincides , modulo 1,
with the one of the composite soliton $(j,j+n-1)$.
A generic elastic plaquette
$\left( \begin{array}{@{\,}c@{\,}c@{\,}} i & l \\ j & k  \end{array} \right)$
( for simplicity we consider all multiplicities equal to one)
describes an elastic scattering process where the "in" state is
given by the ordered set of vacua $(i,j,k)$ and the "out" state by
the set $(i,l,k)$. The $w$-coordinates of the "in" state are the two
circular angles $2  \psi_1$,$2  \psi_2$
determined by the two
incoming solitons.
By elasticity the angles characterizing
the "out" state will be $2  \psi_2$,$2  \psi_1$ (see figure 5).

\begin{figure}
\begin{center}

\unitlength = 1mm
\begin{picture}(110,30)(0,0)
\drawline(5,15)(15,5)(45,15)
\dashline{2}(5,15)(25,25)(15,5)
\dashline{2}(25,25)(45,15)
\drawline(65,15)(95,5)(105,15)
\dashline{2}(65,15)(85,25)(95,5)
\dashline{2}(85,25)(105,15)
\put(55,15){\makebox(0,0){$ \rightarrow$ } }
\put(15,15){\makebox(0,0){$2 \psi_1$}}
\put(27.5,16.5){\makebox(0,0){$2 \psi_2$}}
\put(95,15){\makebox(0,0){$2 \psi_1$}}
\put(82.5,16.5){\makebox(0,0){$2 \psi_2$}}
\put(5,15){\circle*{3}}
\put(15,5){\circle*{3}}
\put(45,15){\circle*{3}}
\put(65,15){\circle*{3}}
\put(95,5){\circle*{3}}
\put(105,15){\circle*{3}}
\put(1,12){\makebox(0,0){$w_i$}}
\put(61,12){\makebox(0,0){$w_i$}}
\put(14,1){\makebox(0,0){$w_j$}}
\put(96,1){\makebox(0,0){$w_l$}}
\put(50,10){\makebox(0,0){$w_k$}}
\put(110,10){\makebox(0,0){$w_k$}}
\end{picture}

\end{center}
\caption[]{Picture on the $w$ plane of an elastic scattering
process }
\end{figure}
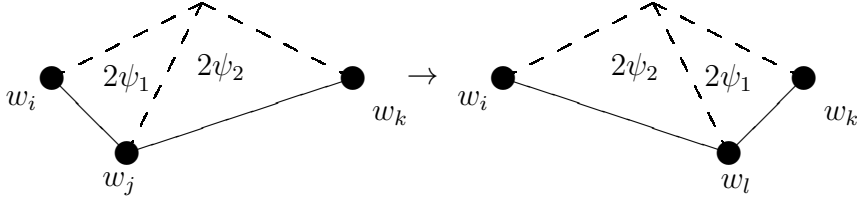

Notice that
the condition of elasticity
allows plaquettes
of the type
$\left( \begin{array}{@{\,}c@{\,}c@{\,}} i & j \\ j & i  \end{array} \right)$
which correspond to the same in and out state with the angles
given by
$\psi_1$ and $-\psi_1$. These type of plaquettes can be
avoided if we impose, in addition to elasticity, the extra condition:

\begin{eqnarray}
f_{i, j} = f_{l, k} &, & f_{j, k} = f_{i, l} \label{11}
\end{eqnarray}

\noindent
i.e. equal fermion number for opposite sides of the plaquette.
As it would be clear from the explicit computations below
,integrability for circular graphs, i.e. solutions to the
gYB equation, will require to impose condition (\ref{11})
on the fermion numbers. This is the type of integrability
symmetries we have discussed in the previous section.
In table 2 we show the ratios between the entries of the
$N=2$ $S-$matrix for the scattering
process described in figure 5.
They can be deduced solving the intertwiner
condition (\ref{8}). The fermion numbers of the incoming
solitons are denoted by $f_1$,$f_2$.

\begin{table}
\begin{center}
\begin{tabular}{|c|c|}
\hline
$\begin{array}{cc} m_3 & m_4 \\ m_1 & m_2 \end{array}$ &
$S^{m_3 \; m_4}_{m_1 \; m_2}( \psi_1 ,\psi_2) /
S^{0 \; 0 }_{0 \; 0} ( \psi_1, \psi_2) $ \\ \hline
$\begin{array}{cc} 1 & 1   \\ 1 & 1 \end{array}$ &
$- e^{{\rm i }  \pi (f_1-  f_2)}
e^{ {\rm i}(\psi_2 - \psi_1)} \frac{ {\rm sinh}( \frac{\theta}{2}
- {\rm i} \frac{ \psi_1+  \psi_2}{2}) }{
{\rm sinh}( \frac{\theta}{2}
+ {\rm i} \frac{ \psi_1+  \psi_2}{2})} $ \\ \hline
$\begin{array}{cc} 1 & 0   \\ 0 & 1 \end{array}$ &
$ e^{{\rm i }  \pi f_1 }
e^{ -{\rm i}  \psi_1} \frac{ {\rm sinh}( \frac{\theta}{2}
- {\rm i} \frac{ \psi_1-  \psi_2}{2}) }{
{\rm sinh}( \frac{\theta}{2}
+ {\rm i} \frac{ \psi_1+  \psi_2}{2})} $ \\ \hline
$\begin{array}{cc} 0 & 1   \\ 1 & 0 \end{array}$ &
$ e^{-{\rm i }  \pi f_2 }
e^{ {\rm i}  \psi_2} \frac{ {\rm sinh}( \frac{\theta}{2}
+ {\rm i} \frac{ \psi_1-  \psi_2}{2}) }{
{\rm sinh}( \frac{\theta}{2}
+ {\rm i} \frac{ \psi_1+  \psi_2}{2})} $ \\ \hline
$\begin{array}{cc} 0 & 1   \\ 0 & 1 \end{array}$ &
$ {\rm i} e^{{\rm i }  \pi (f_1 - f_2) }
e^{- {\rm i} \frac{\psi_1-  \psi_2}{2}}
\frac{ ( {\rm sin}\psi_1  {\rm sin}\psi_2)^{ 1/2 }
 }{{\rm sinh}( \frac{\theta}{2}
+ {\rm i} \frac{ \psi_1+  \psi_2}{2})} $ \\ \hline
$\begin{array}{cc} 1 & 0   \\ 1 & 0 \end{array}$ &
$ {\rm i} e^{- {\rm i} \frac{\psi_1-  \psi_2}{2}}
\frac{ ( {\rm sin}\psi_1  {\rm sin}\psi_2)^{ 1/2 }
 }{{\rm sinh}( \frac{\theta}{2}
+ {\rm i} \frac{ \psi_1+  \psi_2}{2})} $ \\ \hline
 \end{tabular}
\caption{Relations imposed by the N=2 intertwiner
condition on "circular" $S-$matrices. }
\end{center}
\label{}
\end{table}

Once we have obtained the $N=2$ ratios we move into the question
of integrability, i.e. to solve the gYB equation.
Restricting ourselves to elastic plaquettes satisfying (\ref{11}), the
gYB equation reduces to the standard vertex YB equation
where each line is associated with a rapidity $\theta$
(elasticity), "angle" $\psi$
and  fermion number $f$ (condition (\ref{11}). It is not
difficult to check that the $S-$matrix given in table 2 does satisfy
this vertex  Yang-Baxter equation ( there is of course an
overall factor multiplying each $S-$matrix which does not affect
the YB equation).

The solution given in table 2 have a very interesting quantum group
meaning, in fact it is the quantum $R$-matrix intertwiner for the
Hopf algebra $\tilde{U}_{q}(A^{(1)}_1)$ with deformation parameter
$q^4=1$. This Hopf algebra, which was introduced in reference
\cite{Ji}, is very close to the Hopf algebra $U_q(A^{(1)}_1)$. The
only changes are the addition of  two new central elements
$Z_i (i=0,1)$ which modify the usual
comultiplication rules of the rest of the elements
of $U_q(A^{(1)}_1)$ as follows  :

\begin{eqnarray}
& \Delta E_i = E_i \otimes {\bf 1} + Z_i \; K_i \otimes E_i &
\nonumber \\
& \Delta F_i = F_i \otimes K_i^{-1} + Z^{-1}_i \otimes F_i &
\label{12} \\
& \Delta K_i = K_i \otimes K_i & \nonumber \\
& \Delta Z_i = Z_i \otimes Z_i & \nonumber
\end{eqnarray}

We shall be interested in the so called nilpotent representations
\cite{GS2}
of the $\tilde{U}_q(A^{(1)}_1)$  algebra which are labelled,
in addition to the rapidity $\theta$, by
a pair $\xi = (\lambda, z)$ of non  zero complex numbers. They are
given by:

\begin{eqnarray}
\pi_{\l,z}(\theta)(E_0) = d(\l) \left( \begin{array}{cc}
0 & 0 \\  e^{\theta/2}  & 0 \end{array} \right),&
\pi_{\l,z}(\theta)(E_1)= d(\l) \left( \begin{array}{cc}
0 &   e^{\theta/2} \\ 0 & 0  \end{array} \right)&
\nonumber \\
\pi_{\l,z}(\theta)(F_0) = d(\l) \left( \begin{array}{cc}
0 &  e^{-\theta/2} \\ 0  & 0 \end{array} \right),&
\pi_{\l,z}(\theta)(F_1)= \left( \begin{array}{cc}
0 & 0 \\  e^{-\theta/2}  & 0  \end{array} \right)
\label{13}  \\
\pi_{\l,z}(\theta)(K_0) = \left( \begin{array}{cc}
\l^{-1} & 0 \\ 0 &- \l^{-1} \end{array} \right), &
\pi_{\l,z}(\theta)(K_1) = \left( \begin{array}{cc}
\l & 0 \\ 0 & - \l \end{array} \right), & \nonumber \\
\pi_{\l,z}(\theta)(Z_0) = \left( \begin{array}{cc}
z^{-1} & 0 \\ 0 & z^{-1} \end{array} \right), &
\pi_{\l,z}(\theta)(Z_1) = \left( \begin{array}{cc}
z & 0 \\ 0 &  z \end{array} \right), & \nonumber
\end{eqnarray}

\noindent
where $d(\l) = \left( \frac{ \l - \l^{-1}}{2 {\rm i}} \right)^{1/2}$.

The intertwiner matrix $R(\l_1,z_1; \l_2,z_2;\theta)$
of the tensor product of two nilpotents irreps can be computed using
equations (\ref{12}, \ref{13})  and reads:

\begin{eqnarray}
& R^{0 \; 0}_{0 \; 0} = \frac{1}{2} \left( \frac{z_2}{z_1} \right)^{1/2}
\left[ e^{ \theta/2 } ( \l_1 \l_2 )^{1/2}
- e^{- \theta/2 } ( \l_1 \l_2 )^{-1/2} \right] & \nonumber \\
& R^{1 \; 1}_{1 \; 1} =  \frac{1}{2} \left( \frac{z_1}{z_2} \right)^{1/2}
\left[ e^{- \theta/2 } ( \l_1 \l_2 )^{1/2}
- e^{ \theta/2 } ( \l_1 \l_2 )^{-1/2} \right] & \nonumber \\
& R^{1 \; 0}_{0 \; 1} =  \frac{1}{2} \left( z_1 z_2 \right)^{1/2}
\left[ e^{ \theta/2 } \left( \frac{ \l_2}{ \l_1} \right)^{1/2}
- e^{- \theta/2 } \left( \frac{\l_1}{ \l_2} \right)^{1/2} \right] &
\label{14}  \\
& R^{0 \; 1}_{1 \; 0} =  \frac{1}{2} \left( z_1 z_2 \right)^{-1/2}
\left[ e^{ \theta/2 } \left( \frac{ \l_1}{ \l_2} \right)^{1/2}
- e^{- \theta/2 } \left( \frac{\l_2}{ \l_1}
\right)^{1/2} \right] & \nonumber \\
&R^{0 \;1}_{0 \; 1} =  \frac{1}{2}
\left( \frac{ z_1 \l_1}{ z_2 \l_2} \right)^{1/2}
\left[ ( \l_1 - \l_1^{-1})\;  (\l_2 - \l^{-1}_2) \right]^{1/2} &
\nonumber \\
&R^{1 \;0}_{1 \; 0} =  \frac{1}{2}
\left( \frac{ z_2 \l_2}{ z_1 \l_1} \right)^{1/2}
\left[ ( \l_1 - \l_1^{-1})\;  (\l_2 - \l^{-1}_2) \right]^{1/2} &
\nonumber
\end{eqnarray}

The restriction to elastic
plaquettes satisfying the fermion number condition (\ref{11}) allow us
to associate with these plaquettes two irreps $\xi_{1}=(\lambda_{1},z_1)$
and
$\xi_{2}=(\lambda_{2},z_2)$ of $\tilde{U}_{q}(A^{(1)}_1)$
where $\l$ and $z$ are given, in terms of the $N=2$
data, by the following relations:

\begin{eqnarray}
\l =e^{ {\rm i} \psi }, & z = e^{ {\rm i} \pi f}
e^{ - {\rm i } \psi } &
\label{15}
\end{eqnarray}

Using these identifications it is easy to see that the $N=2$ $ S-$matrix
of table 2 coincides, up to an overall factor, with the $R-$matrix (\ref{14}).

Taking into account eq.(\ref{15}) we deduce that the well known
relation between the $N=2$ algebra with the quantum Hopf algebra
$U_q(A^{(1)}_1)$ given by \cite{LVafa,Fe3} :

\begin{eqnarray}
Q^+ = E_1 ,& \bar{Q}^+ = F_1 \; K_1 \label{16} \\
Q^- = E_0 , & \bar{Q}^- = K_0 \; F_0 \nonumber
\end{eqnarray}

\noindent
is completed in the $\tilde{U}_q(A^{(1)}_1)$ algebra by the relation:

\begin{equation}
e^{ {\rm i } \pi {\cal F} } = Z_1 \; K_1
\label{17}
\end{equation}

In summary we obtain the following general result:

{\bf All
${\bf N=2}$ massive theories with all critical points uniformally
distributed on a circle are integrable. Moreover the ${\bf N=2}$
piece of the scattering ${\bf S}$-matrix is given by the quantum
intertwiner of ${\bf \tilde{U}_{q}(A^{(1)}_1)}$ for the irreps defined by
equation (\ref{15}).}

Next we present some Landau-Ginzburg examples:

1) $A_n(t_1): \; \l_a = e^{ \i \pi a/n} , \; z_a=1 \;\;
(a = 1, \dots, n-1)$

2) $D_n(\tau):  \; \l_a = e^{ \i \pi a/n} , \;
z_a= e^{-3 \i \pi a/n} \;\;
(a = 1, \dots, n-1)$.

3) $A_n(t_{n-1}): \; \l = e^{ \i \pi /2} , \; z=1 $

In  cases 1) and 2) the full $S-$matrices are given by the quantum
intertwiners of $\tilde{U}_q(A^{(1)}_1)$.
The only difference
between these two cases resides
in the values of the central elements $z's$ which reflect
the differences in fermion numbers (see eq. (\ref{15}).

The  case 3) corresponds to the Chebishev
potential for the $A$ models. In this case the graph defined by the
vacua coincides with the Coxeter diagram of type $A$. The
$w$-coordinates are degenerate and the image of the graph in the
$w$-plane corresponds to the limit case of the circular graph
defined by two points linked by a line. The $N=2$ piece of the
$S$-matrix is simply the  $R-$matrix of the quantum group
${U}_{q}(A^{(1)}_1)$ for the spin $\frac{1}{2}$ representation.
This is also the $S-$matrix of the sine-Gordon model at coupling
$\beta^2 = \frac{2}{3}
8 \pi$. The $N$=0 piece is given by the ABF
solution for type $A$ $RSOS$ models.

\subsection{Daisy Graphs}

Now we consider uniform daisy graphs of the type depicted in
figure 4,  with $k$
nodes on the circle. We will take as fundamental the radial soliton and
antisolitons $(a,*)$ , $(*,a)$ assigning to both the same
fermion number $f$ which in all
our examples turns out to
be equal to $\frac{1}{2}$
(see table 1). As $w$-coordinates we shall choose zero for the
central node $*$ and $w_a = e^{4 \pi \i a/k} (a=1,\dots,k$)
(this choice is motivated by the
study of the case $A_{k+1}(t_2)$,see table 1).
For these graphs there exist two different types of elastic
plaquettes namely:$\left(\begin{array}{@{\,}c@{\,}c@{\,}} a  & {*}
 \\ {*}   & b   \end{array}\right)$
and$\left( \begin{array}{@{\,}c@{\,}c@{\,}} {*}  & b   \\ a   & {*}
\end{array}\right)$.
As we showed for the circular case ,the $N=2$ invariance fixes the
ratios between the various entries of the daisy $S-$matrices. In table
3 we collect our results for the two types of plaquettes.

\begin{table}
\begin{center}
\begin{tabular}{|c|c|c|}
\hline
$\begin{array}{@{\,}c@{\,}c@{\,}} k & \ell \\ i & j \end{array}$ &
$S^{k \; \ell}_{i \; j} \left(
\begin{array}{@{\,}c@{\,}c@{\,}} a  & {*}   \\ {*}   & b   \end{array}
\right) /
S^{0 \; 0 }_{0 \; 0} \left(
\begin{array}{@{\,}c@{\,}c@{\,}} a  & {*}   \\ {*}   & b   \end{array}
\right)$ &
$S^{k \; \ell}_{i \; j} \left(
\begin{array}{@{\,}c@{\,}c@{\,}} {*}  & {b}   \\ {a}   & {*}   \end{array}
\right) /
S^{0 \; 0 }_{0 \; 0} \left(
\begin{array}{@{\,}c@{\,}c@{\,}}  {*} & b  \\ a &  {*}     \end{array}
\right)$ \\ \hline
$\begin{array}{@{\,}c@{\,}c@{\,}} 1 & 1   \\ 1 & 1 \end{array}$ &
$\frac{cosh( \frac{\theta}{2}+\frac{2 \pi {\rm i} n}{k})
}{cosh( \frac{\theta}{2}-\frac{2 \pi {\rm i} n}{k}) }$  &
$exp(-\frac{4 \pi {\rm i} n}{k})$ \\ \hline
$\begin{array}{@{\,}c@{\,}c@{\,}} 1 & 0    \\ 0 & 1 \end{array}$ &
$-\frac{{\rm i} exp(2 \pi {\rm i} n/k) sinh(\frac{\theta}{2})
}{cosh( \frac{\theta}{2}-\frac{2 \pi {\rm i} n}{k}) } $ &
$-\frac{{\rm i} exp(-2 \pi {\rm i} n/k)
sinh(\frac{\theta}{2} -\frac{2 \pi {\rm i} n}{k})
}{cosh \frac{\theta}{2} }$ \\ \hline
$\begin{array}{@{\,}c@{\,}c@{\,}} 0 & 1    \\ 1 & 0 \end{array}$ &
$-\frac{  {\rm i} exp(-2 \pi {\rm i} n/k) sinh(\frac{\theta}{2})
}{cosh( \frac{\theta}{2}-\frac{2 \pi {\rm i} n}{k}) } $ &
$-\frac{{\rm i} exp(-2 \pi {\rm i} n/k)
sinh(\frac{\theta}{2} +\frac{2 \pi {\rm i} n}{k})
}{cosh \frac{\theta}{2} }$ \\ \hline
$\begin{array}{@{\,}c@{\,}c@{\,}} 1 & 0    \\ 1 & 0 \end{array}$ &
$\frac{cos( \frac{2 \pi n}{k})
}{cosh( \frac{\theta}{2}-\frac{2 \pi {\rm i} n}{k}) }$  &
$\frac{exp(-2 \pi {\rm i} n/k) cos(\frac{2 \pi n}{k} )
}{cosh \frac{\theta}{2} }$  \\ \hline
$\begin{array}{@{\,}c@{\,}c@{\,}} 0 & 1    \\ 0 & 1 \end{array}$ &
$\frac{cos( \frac{2 \pi n}{k})
}{cosh( \frac{\theta}{2}-\frac{2 \pi {\rm i} n}{k}) } $ &
$\frac{exp(-2 \pi {\rm i} n/k) cos(\frac{2 \pi n}{k} )
}{cosh \frac{\theta}{2} }$ \\ \hline
\end{tabular}
\caption{Relations imposed by the N=2 intertwiner
condition}
\end{center}
\label{}
\end{table}

Summing over all elastic plaquettes the gYB equation for daisy
graps is given by:

\begin{eqnarray}
R(\theta_1,\theta_2,\theta_3) \sum_{m'_1,m'_2,m'_3}
S^{m'_1 m'_2}_{m_1 m_2} \left( \begin{array}{@{\,}c@{\,}c@{\,}}
a & {*} \\ {*} & b \end{array}
\right)(\theta_{12})
S^{m'_3 m''_3}_{m'_2 m_3}
\left( \begin{array}{@{\,}c@{\,}c@{\,}} {*}  & c \\ b & {*} \end{array}
\right)(\theta_{13})
S^{m''_1 m''_2}_{m'_1 m'_3}
\left( \begin{array}{@{\,}c@{\,}c@{\,}} a & * \\ {*} & c \end{array}
\right)(\theta_{23})&  & \nonumber \\
= \sum_d \sum_{m'_1,m'_2,m'_3}
S^{m'_2 m'_3}_{m_2 m_3} \left( \begin{array}{@{\,}c@{\,}c@{\,}} {*} & d \\
b & {*} \end{array}
\right)(\theta_{23})
S^{m''_1 m'_1}_{m_1 m'_2}
\left( \begin{array}{@{\,}c@{\,}c@{\,}} a  & {*} \\ {*} & d \end{array}
\right)(\theta_{13})
S^{m''_2 m''_3}_{m'_1 m'_3}
\left( \begin{array}{@{\,}c@{\,}c@{\,}} {*} & c \\ d & {*} \end{array}
\right)(\theta_{12})  & &
\label{18}
\end{eqnarray}

The graphical meaning of this equation is given in figure 6.

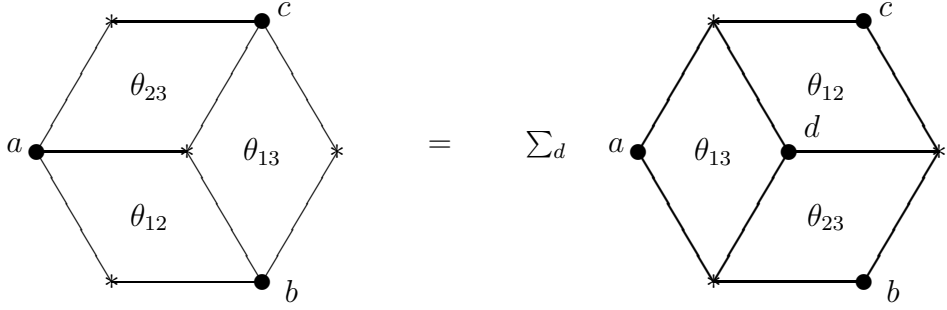
\begin{figure}
\begin{center}

\unitlength = 1mm
\begin{picture}(140,40)(-30,-20)
\drawline(0,0)(10,17.32)(-10,17.32)(-20,0)(0,0)(10,-17.32)
(-10,-17.32)(-20,0)
\drawline(10,17.32)(20,0)(10,-17.32)
\put(-5,-8.7){\makebox(0,0){$\theta_{12}$}}
\put(-5,8.7){\makebox(0,0){$\theta_{23}$}}
\put(10,0){\makebox(0,0){$\theta_{13}$}}
\put(70,0){\makebox(0,0){$\theta_{13}$}}
\put(85,-8.7){\makebox(0,0){$\theta_{23}$}}
\put(85,8.7){\makebox(0,0){$\theta_{12}$}}
\put(0,0){\makebox(0,0){$*$}}
\put(10,17.32){\circle*{2}}
\put(-10,17.32){\makebox(0,0){$*$}}
\put(-20,0){\circle*{2}}
\put(-10,-17.32){\makebox(0,0){$*$}}
\put(10,-17.32){\circle*{2}}
\put(20,0){\makebox(0,0){$*$}}
\thicklines
\drawline(80,0)(100,0)(90,17.32)(70,17.32)(80,0)(70,-17.32)
(90,-17.32)(100,0)
\drawline(70,17.32)(60,0)(70,-17.32)
\put(80,0){\circle*{2}}
\put(100,0){\makebox(0,0){$*$}}
\put(90,17.32){\circle*{2}}
\put(70,17.32){\makebox(0,0){$*$}}
\put(60,0){\circle*{2}}
\put(70,-17.32){\makebox(0,0){$*$}}
\put(90,-17.32){\circle*{2}}
\put(-24,0){$a$}
\put(56,0){$a$}
\put(13,-20){$b$}
\put(93,-20){$b$}
\put(12,18){$c$}
\put(92,18){$c$}
\put(82,2){$d$}
\put(32,0){$=$}
\put(45,0){$\sum_{d}$}
\end{picture}

\end{center}
\caption[]{Star-triangular relation}
\end{figure}

Before entering into the description of the solution to equation
(\ref{18}), we would like to make some preliminary comments on its
structure and physical meaning:

1) The first relevant thing to be noticed is that due to the
geometry of daisy graphs, the gYB equation is asymmetric,
appearing the sum over graph labels only in one of the terms of
the equation. This is a well known phenomena in the $RSOS$
description of chiral Potts models \cite{CP1,CP2} . In fact if we consider
equation (\ref{18}) for the susy labels $m_1$,$m_2$,$m_3$,$m_4$=
$0$,or $1$, the two resulting equations are exactly the RSOS version of
the YB equation for a $Z_k$ chiral Potts model \cite{CP2}.

2) In equation (\ref{18}) we have included an extra factor $R$. The
reason for including this factor is certainly inspired by the
form of the star triangle relation of the
chiral Potts model. As it happens overthere this
factor does not affect the integrability of the theory, which
is certainly the topic
of this section. From a
physical point of view this factor
should be put equal to one
in order to interpret the solutions to (\ref{18}) as the physical
scattering $S$ matrices. We shall follow instead the strategy of
leaving this factor undetermined in the general
discussion  of integrability, while  fixing it in the
scattering theory section. This will require  to
discover the physical spectrum of asymptotic particles.

After these general remarks we are ready to
give the explicit solution. Using the $N=2$ relations of table 3,
the solution is completely characterized by the values of two
entries. After a lengthly computation we find the following solution:

\begin{eqnarray}
S^{0 \; 0 }_{0 \; 0} \left(
\begin{array}{@{\,}c@{\,}c@{\,}} a  & {*}   \\ {*}   & b   \end{array}
\right)(\theta) &=&
S(\theta) \prod^{a-b-1}_{r=0}
\frac{ {\rm cosh}( \frac{\theta}{2} + \frac{2 \pi {\rm i} r}{k}) }{
{\rm cosh} ( \frac{\theta}{2} - \frac{2 \pi {\rm i} r}{k})} \label{19} \\
S^{0 \; 0 }_{0 \; 0} \left(
\begin{array}{@{\,}c@{\,}c@{\,}} {*}  & {b}   \\ {a}   & {*}   \end{array}
\right)(\theta) &=&
\bar{S}(\theta)
e^{2 \pi {\rm i} (a-b)/k} \;\prod^{a-b-1}_{r=0}
\frac{ {\rm sinh}( \frac{\theta}{2} - \frac{2 \pi {\rm i} r}{k}) }{
{\rm sinh} ( \frac{\theta}{2} + \frac{2 \pi {\rm i} (r+1)}{k})} \nonumber
\end{eqnarray}

\noindent
This is a solution to equation (\ref{18}) with the factor $R$ given in
term of the two undetermined functions $S({\theta})$ and
$\bar{S}({\theta})$ as follows:

\begin{eqnarray}
R(\theta_1,\theta_2, \theta_3)& = & \frac{ f(\theta_{12}) f(\theta_{23})}{
f(\theta_{13})}
\nonumber \\
f(\theta)&  = & \frac{ \bar{S}(\theta)}{S(\theta)}
\sum^k_{m=1} e^{ 2 \pi {\rm i} m/k}
\;\; \prod^{m-1}_{r=0}
\frac{ {\rm sinh}( \frac{\theta}{2} - \frac{2 \pi {\rm i} r}{k}) }{
{\rm sinh} ( \frac{\theta}{2} + \frac{2 \pi {\rm i} (r+1)}{k})} \label{20}
\end{eqnarray}

The last formula for the factor $f(\t)$ can be simplified using
the identity:

\begin{equation}
\sum^k_{m=1} e^{ 2 \pi {\rm i} m/k}
\;\; \prod^{m-1}_{r=0}
\frac{ {\rm sinh}( \frac{\theta}{2} - \frac{2 \pi {\rm i} r}{k}) }{
{\rm sinh} ( \frac{\theta}{2} + \frac{2 \pi {\rm i} (r+1)}{k})}
= \sqrt{k} e^{3 \pi \i (k-1)/4}
\prod^{(k-1)/2}_{j=1} \; \frac{ \c \left( \frac{\t}{2} - \frac{
2 \pi \i j }{k} \right) }{  \s \left( \frac{\t}{2} + \frac{
2 \pi \i j }{k} \right)}
\label{21}
\end{equation}

\noindent
The previous solution (\ref{19}) have an intrinsic meaning from the point
of view of chiral Potts. In fact the entries $S^{0\;0}_{0\;0}$
and $S^{1 1}_{1 1}$ are two different trigonometric solution of
the chiral Potts model characterized by a parameter $\omega=
e^{ 4 \pi \i/k}$ , "moduli" $k'=1$ and "chiral angles":

\begin{equation}
\phi = \frac{ \pi }{ 2 } (k \mp 2),\;\;
\bar{\phi} = \frac{ \pi }{ 2 }  (k \pm 2)\;\;
{\rm for}\;\;S^{0 0}_{0 0} \;\; (S^{1 1}_{1 1})
\label{22}
\end{equation}

\noindent

Eq (\ref{22}) suggest some kind of relation with the so called
superintegrable chiral Potts model which corresponds to the
values $\phi= \bar{\phi} = \frac{\pi}{2}$ \cite{Ri}.
This model
has received much attention in the past due to its peculiar properties
which singularize it among the more general class of chiral
Potts models \cite{McCoy}.
The general daisy graph solution we have described above can
be used to analize the integrability of some Landau-Ginzburg
potentials. Next we describe five different models all of them
associated with daisy graphs ( see table 1 for notations).

1) $A_{k+1}(t_2) $, $k$=odd. This model corresponds exactly to the
solution we have just described.

2) $A_{k+1}(t_2) $, $k$=even. In this case we find the phenomena of
degenerate $w-$coordinates as can be
seen from the fact that the $k$ critical
points $x_j = e^{2 \pi \i j/k} (j= 1, \dots, k)$
are mapped onto $k/2$ distinct points
$W_j =e^{4 \pi \i j/k}$ in the $w-$plane.
We can differenciate two different
subcases depending on the parity of $k/2$. If
$k/2$ is even there are three collinear points in the $w$-plane,
while if $k/2$ is odd this situation does not occur.
In the
later case the $N=2$ part of the $S-$matrices is again given
by eqs.(\ref{19}), which must be supplemented with a $N=0$ piece
to take care of the double degeneracy of the vacua in the $w-$plane.
The case of collinear vacua is more subtle since, as can be seen
from table 3, one gets singular values for some entries
of the $S-$matrices. These singularities are due to the fact
that whenever we have three collinear points a two soliton state
is undistinguishable from a single soliton state (see
reference \cite{Ho}).

3) $E_6(t_7)$. The solution of this model is the same as the one
for $A_6(t_2)$.

4) $E_8(t_{16})$.The solution of this model is the same as the one
for $A_8(t_2)$.

5) $D_{k +2 }(t_2)$. For simplicity we consider the $k$= odd
case. From table 1 we observe that the graph is actually three
dimensional with $k$ nodes on the $x$ plane, all on the same circle
and two extra nodes ($*_{\a}, \a=1,2$) on the $y$ line. Inspired by the
solution to the daisy models we can conjecture the following
factorization of the $S$-matrices:

\begin{eqnarray}
S^{m_3 m_4}_{m_1 m_2}
\left( \begin{array}{@{\,}c@{\,}c@{\,}} a & {*_{\beta}}
\\ {*_{\alpha}} & b \end{array}
\right)(\theta) = &
S^{m_3 m_4}_{m_1 m_2} \left( \begin{array}{@{\,}c@{\,}c@{\,}}
a & {*} \\ {*} & b \end{array}
\right)(\theta) \; \;
S \left( \begin{array}{@{\,}c@{\,}c@{\,}} \diamond & \beta \\ \alpha &
\diamond \end{array}
\right)(\theta) & \label{diamond}
\end{eqnarray}

\noindent
and similarly for the other type of plaquettes.
In this factorization the $N=2$ part is the solution for the
daisy graph obtained by proyecting on the $w$-plane. The $N=0$
part is the standard solution for the Ising model
at criticality  where now the
lattice variables are identified with the labels $\a, \beta = 1,2$.

\section{ Scattering Theory}

The question of integrability of a massive $N=2$ theory and its
reduction to a scattering theory satisfying bootstrap and
factorization are two different but related questions. In the
study of integrability we start with a graph and formally assume
that all links of the graph correspond to real asymptotic
particles. In the spirit of this assumption we solve the gYB
equation and interpret the solutions as scattering $S$ matrices.
To promote these solutions, whose existence already implies an
infinite number of conserved charges, to a real scattering $S$
matrix, requires to impose unitarity, crossing and bootstrap.
Only after fulfilling these physical requirements we can be sure
that the $N=2$ massive theory is equivalent to a scattering
theory satisfying Zamolodchikov's axiomatics \cite{ZZ}
and that the links
of the graph actually represent the real asymptotic particles.

In this section we will define a closed, in the bootstrap sense,
scattering theory for the two general types of $N=2$ massive
theories we have described until now, namely those associated with
circular and daisy graphs. The results we find are the
following. For circular graphs the scattering theory is obtained
by solving bootstrap equations which are analogous to the ones
describing Toda type theories \cite{Toda,To1,To2} .
The case of daisy graphs is
physically more interesting, as can be already expected from
the chiral Potts solution. A consistent scattering theory can be
defined only after reducing the physical spectrum to composite
solitons obtained as soliton-antisoliton bound states. The
scattering of these composite solitons is derived from the
chiral Potts solution for radial soliton- antisoliton scattering
by a "fusion" procedure. The resulting scattering theory is
again Toda like of the same type that for the $D_n(\tau)$
model, in the sense that
the central elements $Z_1$
of $\tilde{U}_q(A^{(1)}_1)$ take non
trivial values.  After these introductory remarks we pass to present our
results.

\subsection{ Circular Scattering: Toda like spectrum}

In the circular case the $S-$matrix is given by
the $R-$matrix (\ref{14}) up to an overall factor $Z$
which depends on the irreps $\l_1, \l_2$ which one fixes
imposing unitarity, crossing and bootstrap \cite{Fe2}.
Let us suposse that there are n vacua
equally space on the same circle $j=1,\dots,n$. The value
of $\l$ for the soliton $(j,j+r)$ is given by
$\l = e^{ {\rm i} \pi r/n}$, while we shall leave the value
of $z$ undetermined. Then the $S-$matrix describing the scattering
of the soliton $(j,j+r_1)$ with rapidity $\theta_1$ and the
soliton $(j+r_1,j+r_1+r_2)$ with rapidity $\theta_2$ is given by:

\begin{eqnarray}
&S \left(
\begin{array}{@{\,}c@{\,}c@{\,}} j & j + r_2 \\ j + r_1 & j +r_1+r_2
\end{array}
\right) (\t_{12})=& \nonumber \\
& Z_{r_1, \; r_2}(\t_{12}) \;\;
R ( \l_1= e^{ {\rm i} \pi r_1/n},z_{r_1};
\l_2= e^{ {\rm i} \pi r_2/n},z_{r_2},\t_{12}) &
\label{23}
\end{eqnarray}

Unitarity implies the equation:

\begin{equation}
S \left(
\begin{array}{@{\,}c@{\,}c@{\,}} j & j + r_2 \\ j + r_1 & j +r_1+r_2
\end{array}
\right) (\t) \;\;
S \left(
\begin{array}{@{\,}c@{\,}c@{\,}} j & j + r_1 \\ j + r_2 & j +r_1+r_2
\end{array}
\right) (-\t) = {\bf 1}
\label{24}
\end{equation}

\noindent
and crossing:

\begin{equation}
Z_{r_1, \;r_2}(\t) = Z_{n-r_2, \;r_1}( {\rm i} \pi -\t)
\label{25}
\end{equation}

The analysis of the bootstrap properties of this model leads
finally to the following expresion of $Z_{1 , 1}$ ( see reference
\cite{Fe2} for details):

\begin{eqnarray}
Z_{1, 1 }(\t) =& \frac{1}{ {\rm sinh} \left(
\frac{\t}{2} - \frac{ {\rm i} \pi }{n} \right) }
\prod^{\infty}_{j=1} \;\;
\frac{ \Gamma^2 ( -\t /2 \pi {\rm i} +j)
\Gamma ( \t /2 \pi {\rm i} +j +1/n)
\Gamma ( \t /2 \pi {\rm i} +j -1/n)}{ \Gamma^2 ( \t /2 \pi {\rm i} +j)
\Gamma (- \t /2 \pi {\rm i} +j +1/n)
\Gamma (- \t /2 \pi {\rm i} +j -1/n)} & \label{26} \\
& = \frac{1}{ {\rm sinh} \left(
\frac{\t}{2} - \frac{ {\rm i} \pi }{n} \right) }
exp \left( 2 {\rm i} \int^{\infty}_0 \frac{ {\rm d}t}{t}
{\rm sin} t \t \frac{ {\rm sinh}^2 \pi  t/n }{
{\rm sinh}^2 \pi t }\right) & \nonumber
\end{eqnarray}

We shall return to this factor in the next subsection.

\subsection{ Daisy Scattering: Confinement like spectrum}

The solution (\ref{19}) to the gYB equation for daisy graphs was
fixed up to two undetermined functions $S(\theta)$ and
$\bar{S}(\theta)$. We can  use
this freedom in order to define an unitary and crossing
symmetric S-matrix satisfying at the same time the gYB equation
(\ref{18}) with the factor $R$ set equal to one. It is easy to check
that there is not solution to all these conditions.
Let us show why this happens in more detail.
Crossing symmetry is guaranteed if:

\begin{equation}
S^{m_3 \; m_4}_{m_1 \; m_2} \left(
\begin{array}{@{\,}c@{\,}c@{\,}} {*} & b \\ a & {*} \end{array} \right) (\t) =
S^{ \bar{m}_1 \; m_3}_{m_2 \; \bar{m}_4} \left(
\begin{array}{@{\,}c@{\,}c@{\,}} a & {*} \\ {*} & b \end{array} \right)
( {\rm i} \pi -\t)
\label{27}
\end{equation}

\noindent
where $\bar{m}=1,0$ for $m=0,1$.
This implies in turn
the following relation between $S(\t)$ and $\bar{S}(\t)$:

\begin{equation}
\bar{S}(\t) = {\rm i} \;\;{\rm cotanh} (\frac{\t}{2})
\;\;S( {\rm i} \pi
-\t)
\label{28}
\end{equation}

\noindent
Introducing eq.(\ref{28}) into (\ref{20}) and using the
relation (\ref{21})  one deduces:

\begin{equation}
f(\t) \; f({\rm i} \pi - \t) = k
\label{29}
\end{equation}

\noindent
which already implies that the factor $R$ cannot be set equal to
one in the gYB equation. Moreover, the unitarity conditions:

\begin{eqnarray}
S\left(
\begin{array}{@{\,}c@{\,}c@{\,}} a  & {{*}}   \\ {{*}}   & b   \end{array}
\right)(\theta)
S\left(
\begin{array}{@{\,}c@{\,}c@{\,}} a  & {{*}}   \\ {{*}}   & b   \end{array}
\right)(-\theta)= &  {\bf 1}&
\label{30} \\
\sum_b
S\left(
\begin{array}{@{\,}c@{\,}c@{\,}} {{*}}  & {a}   \\ {b}   & {{*}}   \end{array}
\right)(\theta)
S\left(
\begin{array}{@{\,}c@{\,}c@{\,}} {{*}}  & {b}   \\ {c}   & {{*}}   \end{array}
\right)(-\theta) = & \delta_{a,c} \;\;{\bf 1}  \nonumber
\end{eqnarray}

\noindent
are incompatible with crossing. The best that can be done is to find
solutions which satisfy unitarity, but violate crossing by a
constant and satisfy the gYB equation with the factor $R$ also a
constant. Notice that the violation of crossing and the existence of
the anomalous factor $R$ in the factorization equations, are two
related problems. In fact by crossing we relate the two types of
elastic plaquettes, and the missmatch in the crossing relation
show up in the factor $R$.
The physical origin of these problems can be partially
understood as a consequence of the special symmetry properties
of the chiral Potts Boltzmann weights. In fact the solution
(\ref{19}) is neither $P$ nor $T$ invariant, however it satisfies the
most general requirement of $PCT$ invariance:

\begin{equation}
S^{m_3 \; m_4}_{m_1 \; m_2} \left(
\begin{array}{@{\,}c@{\,}c@{\,}} a & {{*}} \\ {{*}} &  b \end{array} \right)
(\t) =
S^{ \bar{m}_2 \; \bar{m}_1}_{\bar{m}_4 \; \bar{m}_3} \left(
\begin{array}{@{\,}c@{\,}c@{\,}} b & {{*}} \\ {{*}} & a \end{array} \right)
( {\rm i} \pi -\t)
\label{31}
\end{equation}

We shall follow a different strategy
for associating to daisy graphs a well
defined scattering theory without the anomalies described above.
First of all we reduce the spectrum to
composite solitons $(a,b)$ defined by a pair of
radial soliton $(a,{*})$ and antisoliton $({*},b)$
,then by means of a fusion
procedure we compute  their $S$ matrix and
finally we use the freedom of the
unknown functions
$S(\t)$ and $\bar{S}(\t)$ to solve the bootstrap
equations. To simplify matters we shall work out
explicitly the simplest non trivial
case  $k$=3. We proceed in two steps: i) fusion and ii) bootstrap.

{\bf i) Fusion:}

In order to discuss the case $ k=3$ with some detail it is quite
convenient to construct the table 4 which gives the relevant
$S-$matrix elements between the solitons $(a,{*})$ and $({*},b)$.

\begin{table}
\begin{center}
\begin{tabular}{|c|c|c|c|}
\hline
$a-b$ & $0$ & $1$ & $2$ \\ \hline
$S^{0 \; 0}_{0 \; 0}
\left( \begin{array}{@{\,}c@{\,}c@{\,}} a & {*} \\ {*} & b \end{array} \right)
$ & $S(\t)$ & $S(\t)$ & $S(\t)
\frac{ {\rm cosh}\left( \frac{\t}{2} + \frac{ 2 \pi {\rm i} }{3}
\right)}{
{\rm cosh}\left( \frac{\t}{2} - \frac{ 2 \pi {\rm i} }{3} \right)}$
\\ \hline
$S^{1 \; 1}_{1 \; 1}
\left( \begin{array}{@{\,}c@{\,}c@{\,}} a & {*} \\ {*} & b \end{array} \right)
$
&  $S(\t)$ & $S(\t)
\frac{ {\rm cosh}\left( \frac{\t}{2} + \frac{ 2 \pi {\rm i} }{3}
\right)}{
{\rm cosh}\left( \frac{\t}{2} - \frac{ 2 \pi {\rm i} }{3} \right)}$
& $S(\t)$ \\ \hline
$S^{0 \; 0}_{0 \; 0}
\left( \begin{array}{@{\,}c@{\,}c@{\,}} {*} & b \\ a & {*} \end{array} \right)
$
& $\bar{S}(\t)$ & $\bar{S}(\t)
\frac{ e^{2 \pi {\rm i}/3} {\rm sinh} \frac{\t}{2} }
{ {\rm sinh} \left( \frac{\t}{2} + \frac{2 \pi {\rm i}}{3} \right)}$
& $\bar{S}(\t)
\frac{ e^{-2 \pi {\rm i}/3} {\rm sinh} \frac{\t}{2} }
{ {\rm sinh} \left( \frac{\t}{2} + \frac{2 \pi {\rm i}}{3} \right)}$
\\ \hline
$S^{1 \; 1}_{1 \; 1}
\left( \begin{array}{@{\,}c@{\,}c@{\,}} {*} & b \\ a & {*} \end{array} \right)
$
& $\bar{S}(\t)$ & $\bar{S}(\t)
\frac{ e^{-2 \pi {\rm i}/3} {\rm sinh} \frac{\t}{2} }
{ {\rm sinh} \left( \frac{\t}{2} + \frac{2 \pi {\rm i}}{3} \right)}$
& $\bar{S}(\t)
\frac{ e^{2 \pi {\rm i}/3} {\rm sinh} \frac{\t}{2} }
{ {\rm sinh} \left( \frac{\t}{2} + \frac{2 \pi {\rm i}}{3} \right)}$
\\ \hline
 \end{tabular}
\caption{ $ S-$matrix elements for $k=3$ }
\end{center}
\label{}
\end{table}

Looking at the soliton-antisoliton process $s_{a,{*}}(\t_1) +
s_{{*},b}(\t_2)$ one obtains the Bogomolnyi solitons $(a \pm 1 ,a)$
for the value $\t_{12}  = \frac{ \i \pi}{3} $.
If moreover the function $S(\t)$ where non vanishing at
$\t  = \frac{ \i \pi}{3}$  then these states would simply be
bound states of the radial ones. We shall see at the end of this
section what is the fate of $S(\t= \frac{ \i \pi}{3} )$.
In any case and for our
purposes we shall consider the states
$(a \pm 1,a)(\t) $, which have a  mass $m_{a \pm 1 , a}
= 2 \; m_{{*} , a}\; {\rm cos}( \mu/2) \;\; ( \mu = \frac{ \pi}{3})$
, as made of the pairs
$s_{a \pm 1,{*}}(\t + \i \mu/2 ) \otimes
s_{{*},a}(\t - \i \mu/2)$. This will allow us to compute the
$S-$matrix associated with the plaquettes:
$\left(\begin{array}{@{\,}c@{\,}c@{\,}} a \pm 1  & a
 \\ a   & a \mp 1   \end{array}\right)$
by the fusion of four elementary plaquettes (see figure 7):

\begin{eqnarray}
& S\left( \begin{array}{@{\,}c@{\,}c@{\,}} a \pm 1  & a
\\ a   & a \mp 1   \end{array}\right)(\t)=  & \nonumber \\
& \sum_b \; S\left(\begin{array}{@{\,}c@{\,}c@{\,}} a \pm 1  & {*}
\\ {*}   & b   \end{array}\right) (\t) \;
S\left(\begin{array}{@{\,}c@{\,}c@{\,}} b  & {*}
\\ {*}   & a \mp 1   \end{array}\right) (\t)
S\left(\begin{array}{@{\,}c@{\,}c@{\,}} {*}  & b
\\ a   & {*}   \end{array}\right) (\t -\i \mu) \;
S\left(\begin{array}{@{\,}c@{\,}c@{\,}} {*}  & a
\\ b   & {*}   \end{array}\right) (\t + \i \mu)   &
\label{32}
\end{eqnarray}

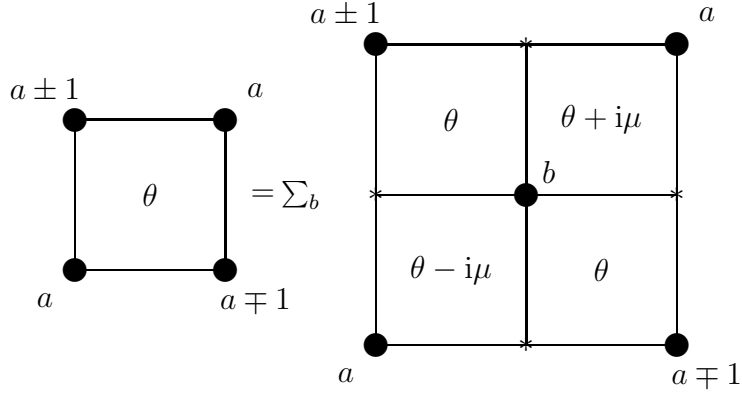
\begin{figure}
\begin{center}

\unitlength = 1mm
\begin{picture}(100,60)(0,-30)
\drawline(10,-10)(30,-10)(30,10)(10,10)(10,-10)
\put(10,-10){\circle*{3}}
\put(30,-10){\circle*{3}}
\put(30,10){\circle*{3}}
\put(10,10){\circle*{3}}
\put(20,0){\makebox(0,0){$\theta$}}
\put(60,10){\makebox(0,0){$\theta$}}
\put(80,-10){\makebox(0,0){$\theta$}}
\put(80,10){\makebox(0,0){$\theta + {\rm i} \mu$}}
\put(60,-10){\makebox(0,0){$\theta-{\rm i} \mu $}}
\put(35,0){\makebox(0,0){$=$}}
\put(40,0){\makebox(0,0){$\sum_b$}}
\put(50,0){\makebox(0,0){$*$}}
\put(90,0){\makebox(0,0){$*$}}
\put(70,20){\makebox(0,0){$*$}}
\put(70,-20){\makebox(0,0){$*$}}
\put(6,14){\makebox(0,0){$ a \pm 1$}}
\put(34,14){\makebox(0,0){$ a $}}
\put(6,-14){\makebox(0,0){$ a $}}
\put(34,-14){\makebox(0,0){$ a \mp 1$}}
\put(46,24){\makebox(0,0){$ a \pm 1$}}
\put(94,24){\makebox(0,0){$ a $}}
\put(46,-24){\makebox(0,0){$ a $}}
\put(94,-24){\makebox(0,0){$ a \mp 1$}}
\put(73,3){\makebox(0,0){$b$}}
\drawline(50,-20)(90,-20)(90,20)(50,20)(50,-20)
\drawline(50,0)(90,0)
\drawline(70,20)(70,-20)
\put(50,-20){\circle*{3}}
\put(90,-20){\circle*{3}}
\put(90,20){\circle*{3}}
\put(50,20){\circle*{3}}
\put(70,0){\circle*{3}}
\end{picture}

\end{center}
\caption[]{Fusion of daisy $S-$matrices to produce circular ones. }
\end{figure}

\noindent
In these equations the $N=2$ labels have been skipped for
simplicity ( they are all 1's or 0's
for the upper or lower choices of signs). Using now
table 4 we obtain :

\begin{eqnarray}
S\left( \begin{array}{@{\,}c@{\,}c@{\,}} a \pm 1  & a
\\ a   & a \mp 1   \end{array} \right) (\t) =
- 3 \i \frac{  \s \frac{\t}{2} \s \left( \frac{ \t}{2} + \i
\frac{\mu }{2} \right) \s \left( \frac{ \t}{2} - \i \mu \right) }{
\c \frac{\t}{2}  {\c}^2 \left( \frac{ \t }{2} - 2 \i \mu \right) }
S(\t)^2 \;\; \bar{S}( \t + \i \mu ) \;\; \bar{S}( \t - \i \mu) &
&
\label{33}
\end{eqnarray}

\noindent
Considering the remaining $N=2$ entries of this
$S-$ matrix one recovers the
intertwiner $R$ matrix of
$\tilde{U}_{q}(A^{(1)}_1)$ Hopf algebra:

\begin{eqnarray}
S\left(\begin{array}{@{\,}c@{\,}c@{\,}} a \pm 1  & a
\\ a   & a \mp 1   \end{array}\right)(\t)=&
3 \i \frac{  \s \frac{\t}{2} \s \left( \frac{ \t}{2} + \i
\frac{\mu}{2} \right) }{
\c \frac{\t}{2}  \c^2 \left( \frac{ \t }{2} - 2 \i \mu \right) }
S(\t)^2 \;\; \bar{S}( \t + \i \mu ) \;\; \bar{S}( \t - \i \mu)
& \nonumber \\
& R \left( \l_1 = \l_2 = \pm e^{ \pm \i \mu },
z_1 =z_2 = e^{\mp \i
\mu }, \t \right) &
\label{34}
\end{eqnarray}

The overall factor which depends on $S(\theta)$ and $\bar{S}(\theta)$
will be fix by bootstrap.
The fused $S$ matrix satisfies the gYB equation for the circular
graph defined by the subset of nodes, of the daisy graph, in
this particular case three, living on the circle. The fusion
procedure we have used able us to get rid of the
unpleasent $R$ factors which obscure the physics of the model.

{\bf{ii) Bootstrap}}

Our next step will be to fix the overall factors by imposing
crossing, unitarity and bootstrap on the fused scattering $S-$
matrix. The geometry of the graph and the
dependence of this overall factor on the undetermined functions
$S(\theta)$ and $\bar{S}(\theta)$  suggest already
the answer, namely the bootstrap factors
of circular graphs. It is rather amusing and
interesting to observe that the structure of the bootstrap
factors $Z_{1, 1 }= Z_{2, 2}$
for circular graphs (\ref{26}) agrees with the one obtained
in (\ref{34}) by fusion, provided we make the following identification:

\begin{eqnarray}
Z_{1, 1}(\t) = - 3 \i  \frac{ \s \frac{ \t}{2}
\c \left( \frac{\t }{2}
+ \frac{ \i \pi}{6} \right)
\c \left( \frac{\t }{2}
- \frac{ \i \pi}{6} \right) }{ \c
\frac{\t }{2}  {\c}^2 \left( \frac{\t }{2}
- \frac{2   \pi \i }{3} \right)
\s \left( \frac{\t}{2}
- \frac{ \i \pi}{6} \right) }
S(\t )^2 \; S( \frac{ 2 \pi \i  }{3} - \t ) \;
S( \frac{ 4 \pi \i }{3} - \t )  & &
\label{35}
\end{eqnarray}

\noindent
{}From the equation (\ref{26}) we finally get the following expresion
of $S(\t)$:

\begin{equation}
S(\t) = C \; \frac{ \c \left( \frac{\t}{2} - \frac{2 \pi \i}{3}
\right) }{ \left( \s \frac{3 \t}{2} \right)^{1/2} }
\prod^{\infty}_{j=1}
\frac{ \Gamma \left( \x +j + \frac{1}{3} \right)
\Gamma \left( \x +j - \frac{1}{3} \right)}{
\Gamma \left(- \x +j + \frac{1}{3} \right)
\Gamma \left(- \x +j - \frac{1}{3} \right)}
\label{36}
\end{equation}

\noindent
where $C$ is some constant.
Strictely speaking eq.(\ref{36}) is formal since the infinite
productorium is actually divergent!. This divergency actually
cancells out in equation (\ref{35}) when reproducing $Z_{1,1}$
from $S(\t)$.
All these considerations provide a certain amount of evidence for
interpreting the radial solitons as elementary constituents and
the physical particles of daisy graph models as composite solitons.
With respect to this second point we should make the following
remark. From relations (\ref{36}) we get the explicit value of the
function  $S(\theta)$  , this function have a zero at the value
$\theta=\frac{i\pi}{3}$ (not taking into
account the divergence mention above) and therefore the composite state
$(a \pm 1 ,a)$ cannot be taken "sensu stricto" as a bound state
of two radial solitons. This can be interpreted
as reflecting some kind of confinement of the  radial
solitons i.e. that they
cannot appear as real asymptotic particles.

To finish this section we would like to make a comment on the
possible scattering theory for the $D_{k+2}(t_2)$ model. Fusing the
tentative solution conjectured in equation (\ref{diamond}) we can obtain
the scattering matrices for a confined
spectrum containing now two different types of composite solitons
namely $(a,*_{\a},b)$
for $\a=1,2$. These $S$ matrices are the product of a $N=0$
part obtained by fusing the Ising components and the same $N=2$
part that for the $A_{k+1}(t_2)$.

The physical picture
that emerges from the previous analysis of the daisy graph
scattering is quite intriging and certainly deserves a deeper
study.

{\bf Acknowledgements}: This work has been supported by the
spanish grant n§ PB92-1092.

\end{document}